\documentstyle[12pt]{article}
\begin{document}
\vspace{-0.2cm}
\begin{flushright}
hep-th/9903032 \\
\vskip2mm
March 1999
\end{flushright}
\vskip.3cm
\begin{center}
{ \bf  SPINNING PARTICLE AS A NON-TRIVIAL ROTATING \\
\vskip3mm
SUPER BLACK HOLE WITH BROKEN N=2 \\
\vskip3mm
SUPERSYMMETRY}\\
\vskip.3cm
{\bf A.Ya. Burinskii} \footnote{ e--mail: grg@ibrae.ac.ru}
\vskip2mm
{\em Gravity Research Group, NSI Russian Academy of Sciences,\\
B.Tulskaya 52, 113191 Moscow, Russia}
\end{center}
\vskip .3cm
\centerline{\bf Abstract}
\par
\begin{quotation}

A non-trivial supergeneralization of the Kerr-Newman solution
is considered as representing a combined model of the Kerr-Newman
spinning particle and superparticle.

We show that the old problem of obtaining non-trivial super black hole
solutions can be resolved in supergravity broken by Goldstone fermion.
Non-linear realization of broken N=2 supersymmetry specific for the
Kerr geometry is considered and some examples of the super-Kerr
geometries generated by Goldstone fermion are analyzed.
The resulting geometries acquire torsion, Rarita-Schwinger field and
extra wave contributions to metric and electromagnetic field caused
by Grassmann variables.

One family of the self-consistent super--Kerr--Newman solutions to
broken N=2 supergravity is selected, and peculiarities of these
solutions are discussed. In particular, the appearance of extra `axial'
singular line and traveling waves concentrated near `axial' and
ring-like singularities.
\end{quotation}
\par
\newpage
\section{Introduction}
\par
 Since 1968 it has been mentioned that the Kerr-Newman solution
possesses some remarkable properties which allow to consider it as a
model of spinning particle \cite{caris,str,bls,c-str}. Some string-like
structures were obtained in Kerr geometry. The first one is connected
with a singular ring of the Kerr solution \cite{str,N-str}, two others are
linked with a complex representation of Kerr geometry \cite{BKP} in which
the Kerr-Newman
solution is considered as a retarded-time field generated by a complex
source propagating along a complex world line \cite{c-str}.
  The point of view that some of black holes can be treated as elementary
particles was also suggested by superstring theory
\cite{part,Sen,BS,c-str,N-str}.
  In particular, Sen \cite{Sen} has obtained a generalization of the Kerr
solution to low energy string theory, and it was shown \cite{BS} that near
the Kerr singular ring the Kerr-Sen solution acquires a metric similar to
the field around a heterotic string.
\par
However, description of spinning particle based only on the bosonic
fields cannot be complete, and involving fermionic degrees of freedom
is required. The most natural way to involve fermions is to treat
corresponding super black holes in supergravity.
In previous papers \cite{super-K,superBH} we considered a method of
supergeneralization of the Kerr-Newman geometry leading to a combined
model of the Kerr spinning particle and superparticle .
It was mentioned that this method leads to the broken or
non-linear realization of supersymmetry.
\par
In the present paper we show that the breaking of supersymmetry caused by
Goldstone fermion allows to resolve the old problem of non-triviality of
the super black holes solutions \cite{striv,finkim,AG1,Aich}.
 Using Deser-Zumino model of broken N=1 supergravity  \cite{DZ}
generalized to the N=2 supergravity described by Ferrara and
Nieuwenhuizen \cite{FN}, we analyze some non-trivial supergeneralizations
of the Kerr-Newman geometry  and show the appearance of torsion and
traveling waves.
\par
We select one self-consistent class of the super-Kerr-Newman solutions
to N=2 supergravity with supersymmetry broken by Goldstone fermion and
discuss peculiarities of these solutions. In particular, the appearance
of extra `axial' singular line and traveling waves concentrated near
the `axial' and ring-like singularities.
\par
\section{ Non-trivial super-solutions from trivial ones}
\par
The problem of non-trivial supergravity solutions is connected
with the fact that any solution of Einstein gravity is indeed a trivial
solution of supergravity field equations with a zero spin-3/2 field
\cite{striv,finkim,AG1,Aich}.
By using a supergauge freedom of supergravity (supertranslations)
one can turn the gravity solutions into a form containing spin-3/2
field. However, since this field can be gauged away by the reverse
transformation, such supersolutions are indeed
{\it trivial}.
There existed even an opinion that all the super black hole solutions
are trivial. However, some examples of the non-trivial super
black hole solutions were given by Aichelburg and G\"uven \cite{AG2,AG3},
and also in two dimensional dilaton supergravity by Knutt-Wehau
and Mann \cite{KnMann}.
\par
In previous papers \cite{super-K,superBH} we showed that
{\it non-trivial} examples of the super-Kerr geometry can be obtained by a
{\it trivial} supershift of
the Kerr solution taking into account some non-linear B-slice constraints.
Indeed, the complex structure of the Kerr geometry
prompts how to avoid this triviality problem.
\par
The Kerr-Schild form of the Kerr geometry \cite{DKS}
\begin{equation}
g_{ik} = \eta_{ik} + 2h k_i k_k
\label{KS}
\end{equation}
allows to give a complex representation of the Kerr solution as a geometry
generated by a complex source  propagating along a complex world line
$x^i_0 (\tau)$ in auxiliary Minkowski space $\eta= diag(-,+++)$.
This representation shows that from complex point of view the Schwarzschild
and Kerr geometries are equivalent and connected by a {\it trivial} complex
shift.
\par
The {\it non-trivial} twisting structure of the Kerr geometry arises as a
result of the {\it shifted real slice} of the complex
retarded-time construction \cite{BKP,c-str}. If the real slice is passing via
`center' of the solution $x_0$ there appears a usual spherical symmetry
of the Schwarzschild geometry. The specific twisting structure
results from the complex shift of the real slice regarding the source.
\par
Similarly, it is possible to turn a {\it trivial} super black hole
solution into a {\it non-trivial} if one finds an analogue to the real slice
 in superspace.
\par
 The {\it trivial supershift} can be represented as a
replacement of the complex world line by a superworldline
\begin{equation}
 X^i_0(\tau)= x^i_0(\tau)-i \theta \sigma ^i \bar \zeta +i\zeta
\sigma^i \bar \theta,
\label{swl} \end{equation}
parametrized by Grassmann coordinates
$\zeta, \quad \bar \zeta$, or as a corresponding coordinate
replacement in the Kerr solution
\begin{equation}
x^{\prime i}  = x^i + i \theta\sigma^i \bar \zeta
 - i \zeta\sigma^i \bar \theta;
\qquad
\theta^{\prime}=\theta + \zeta ,\quad
{\bar\theta}^{\prime}=\bar\theta + \bar\zeta, \label{SG}
\end{equation}
\par
Assuming that coordinates $x^i$ before the supershift are the usual
c-number coordinates one sees that coordinates acquire nilpotent
Grassmann contributions after supertranslations. Therefore, there
appears a natural splitting of the space-time coordinates on the
c-number `body'-part and a nilpotent part - the so called `soul'.
The `body' subspace of superspace, or B-slice, is a submanifold
where the nilpotent part is equal to zero,
and it is a natural analogue to the real slice in complex case.
\par
It has been shown in \cite{super-K,superBH} that reproducing the real slice
procedure of the Kerr geometry in superspace we have to consider
super light cone constraints \footnote{These constraints are similar to
the complex light cone constraints of the standard Kerr geometry
connected with a retarded-time construction. The physical sence of
these constraints is existence of the real slice for the light cones
placed at the points of complex world line $x^i_0(\tau)$. Similarly,
super light cone constraints demand existence of the body-slice for
the super light cones placed at the points of the super world line
(\ref{swl}).}
\begin{equation}
s^2= [x_i-X_{0i}(\tau)][x^i - X^i_0(\tau)] =0 ,
\label{slc}\end{equation}
and B-slice, where coordinates $x^i$ do not contain nilpotent contributions.
Selecting the body and nilpotent parts of this equation we obtain three
equations. The first one is the real slice condition of the complex Kerr
geometry claiming that complex light cones, described by set
\begin{equation}
x=x_0 (\tau ) + \Psi \sigma \tilde \Psi, \label{Psi}
\label{cone}\end{equation}
can reach the real slice.
Here $\Psi$ and $\tilde \Psi$ are the commuting two-component spinors.
On the real slice field  $\Psi (x)$ determines the principal
null congruence of the Kerr geometry (\ref{KS})
\begin{equation}
k_i (x) = P^{-1} \Psi \sigma _i \bar \Psi .\label{k}
\label{lk}\end{equation}
The nilpotent part of (\ref{slc}) yields two B-slice conditions
\begin{equation} [x^i-x_0^i (\tau)]
( \theta\sigma_i \bar \zeta
 - \zeta\sigma_i \bar \theta)=0; \label{odd1}\end{equation}
\begin{equation}
( \theta\sigma \bar \zeta
 - \zeta\sigma \bar \theta)^2 =0.\label{odd2}\end{equation}
Equation (\ref{odd1}) may be rewritten using (\ref{Psi}) in the
form
\begin{equation}
  (\theta^\alpha\sigma_{i\alpha\dot\alpha}\bar\zeta^{\dot\alpha}
 - \zeta^\alpha\sigma_{i\alpha\dot\alpha}\bar\theta^{\dot\alpha})
\Psi^\beta \sigma^i_{\beta\dot\beta} {\bar\Psi}^{\dot\beta}
=0 \label{odd4}\end{equation}
which yields
\begin{equation}
\bar\Psi \bar\theta =0,\qquad\bar\Psi \bar\zeta =0,
\label{odd5}\end{equation}
which in turn is a condition of proportionality of the commuting spinors
$\bar\Psi(x)$ and anticommuting spinors $ \bar\theta$ and $\bar\zeta$,
this condition providing the left  null superplanes of the supercones to
reach B-slice.
\par
Finally, by introducing the Kerr projective spinor coordinate $Y(x)$
we have ${\bar \Psi}^{\dot 2}=Y (x),
\quad{\bar \Psi} ^{\dot 1}=1$, and we obtain
\begin{equation}
{\bar\theta}^{\dot \alpha} =
\left(\begin{array}{c}
 {\bar\theta}^{\dot 1}\\
Y(x){\bar\theta}^{\dot 1}
\end{array} \right),
\label{subman}\end{equation}
\begin{equation}
{\bar\zeta}^{\dot \alpha} =
\left(\begin{array}{c}
 {\bar\zeta}^{\dot 1}\\
Y(x){\bar\zeta}^{\dot 1}
\end{array} \right).
\label{sshift}\end{equation}
\par
It also leads to $ \bar\theta \bar\theta= \bar\zeta \bar\zeta=0, $
and equation (\ref{odd2}) is satisfied automatically.
\par
Thus, as a consequence of the B-slice and superlightcone constraints
we obtain a non-linear submanifold of superspace $\theta = \theta (x),
\quad \bar \theta = \bar \theta (x).$
The original four-dimensional
supersymmetry is broken, and the initial supergauge freedom  which
allowed to turn the super geometry into trivial one is lost. Nevertheless,
there is a residual supersymmetry based on free Grassmann parameters
$ \theta ^1, \quad \bar \theta ^1$.
\par
It was mentioned that the above B-slice constraints yield in fact the
non-linear realization of broken supersymmetry introduced by Volkov
and Akulov \cite{VA,WB} and considered in N=1 supergravity by Deser and
Zumino \cite{DZ}.
In terminology of broken supersymmetry the Grassmann parameters
$\zeta ^\alpha (x), \quad \bar \zeta ^{\dot\alpha} (x) $
represent some fermion Goldstone field on space-time  which has to be eaten
by spin-3/2 Rarita-Schwinger field. As a result supergauge will be fixed.
\par
In this paper we consider the application of this approach to obtaining the
superversions of the Kerr-Newman solution to N=2 supergravity formulated
in \cite{FN}.
\par
\section{Broken Supersymmetry in N=1 Supergravity}
\par
Here we use spinor notations of the book \cite{WB}. The indices
$i,j,k,l...$ are related to coordinate, and $a,b,c,d...$ are reserved
for tetrad.
\par
The Volkov-Akulov model of non-linear realization of supersymmetry
\cite{VA,WB} is based on selecting a submanifold of superspace setting the
correspondence of the Grassmann coordinates
\footnote { Further we use the Dirac four-component spinor notations.}
\begin{equation}
\Theta=
\left(\begin{array}{c}
 {\theta}_{\alpha}\\
{\bar\theta}^{\dot\alpha}
\end{array} \right)
\label{sub1}\end{equation}
to a Goldstone field
$\lambda (x) $ which is a Majorana fermion. This yields
the submanifold $ \Theta (x) = b\lambda (x) $ which is non-linear
in general case.
The non-linear residual supertransformations are
\begin{equation}
\delta _\epsilon \lambda = b^{-1} \epsilon + ib(\bar \epsilon
\gamma ^i \lambda) \partial _i \lambda (x),
\label{nlt}\end{equation}
and contain inhomogeneous term
$b^{-1} \epsilon =
\left(\begin{array}{c}
 {\zeta}_{\alpha}\\
{\bar\zeta} ^{\dot\alpha}
\end{array} \right)$.
\par
Considered by Deser and Zumino case of broken N=1 supergravity
is based on this model \cite{DZ}, and it is proposed that $\epsilon $ admits
local transformations $\epsilon (x)$.
\par
The Lagrangian is given by
\footnote{We have omitted here cosmological term since here we shall
consider only the region of massless fields.}
\begin{equation}
{\cal L} = - ( i / 2 ) \bar \lambda \gamma {\cal D} \lambda - ( i / 2 b )
\bar \lambda \gamma ^i \psi _i + {\cal L} _{sg},
\label{L}\end{equation}
where the supergravity Lagrangian is
\begin{equation}
{\cal L} _{sg}= -eR/2k^2 - i/2 \epsilon^{ijkl} \bar \psi _i \gamma _5
\gamma _j {\cal D} _k \psi _l,
\label{Lsg}\end{equation}
and
\begin{equation}
{\cal D} _i = \partial _i - \frac{1}{2} \omega _{iab} \Sigma ^{ab};
\qquad \Sigma ^{ab} = \frac{1}{4} [\gamma ^a, \gamma ^b ].
\label{D}\end{equation}
The Lagrangian is invariant under the above non-linear
supertransformations, and tetrad $e^a$ and Rarita-Schwinger fields
$\psi_i$ are transformed as follows
\begin{equation}
\delta _\epsilon \lambda =b^{-1} \epsilon + ib(\bar \epsilon
\gamma ^i \lambda) \partial_i \lambda , \label{a}
\end{equation}
\begin{equation}
\delta _\epsilon e^a_i =-ik\bar \epsilon \gamma ^a \psi _i,
\label{b}
\end{equation}
\begin{equation}
\delta _\epsilon \psi _i  =-2/k {\cal D}_i \epsilon . \label{c}
\end{equation}
\par
It is assumed that this construction is similar to the Higgs
mechanism of the usual gauge theories. The Goldstone fermion
$\lambda (x)$ can be eaten by appropriate local transformation
$\epsilon (x)$ with  a corresponding redefinition of the tetrad
and spin-3/2 field. It means that starting from the gravity
solution with zero spin-3/2 field and some Goldstone fermion
field $\lambda$ one can obtain in such a way a non-trivial
supergravity solution with non-linear realization of broken
supersymmetry.
\par
There are  two obstacles for indirect application of this
scheme to the Kerr-Newman case. First one is the electromagnetic charge
which demands to change the expression for supercovariant derivative that
leads to non-Majorana values for spin-3/2 field. The second one is the
complex character of supertranslations in the Kerr case that also
yields the non-Majorana supershifts. Thus, this scheme has to be
extended to N=2 supergravity.
\par
\section{Broken supersymmetry in N=2 supergravity}
\par
The consistent N=2 supergravity described by Ferrara and
Nieuwenhuizen \cite{FN} is based on the complex (non-Majorana)
supergauge field  $\epsilon =(\epsilon _1 + i \epsilon _2)/\sqrt{2}$
 and contains a complex spin-3/2 field
  $\chi =(\psi + i \phi)/\sqrt{2}$ and the vector potential of
  electromagnetic field $A_i$.
\par
The expression for supercovariant derivative (see Appendix B) is extended
by electromagnetic contribution $F_{ab}$ and torsion terms
${\cal D} _{N i} =  (k^2/2)\Sigma ^{ab}k_{iab}$,
\begin{equation}
{\tilde {\cal D}} _{i} = {\cal D} _i +{\cal D} _{Ni} - (ik/2\sqrt{2}) F_{ab}
\Sigma ^{ab}\gamma _i.
\end{equation}
The action is invariant under the following complex local
supersymmetry transformations
\begin{equation}
\delta _\epsilon e^a_i =k(\bar \epsilon \gamma ^a \chi _i -
 \bar \chi _i \gamma ^a \epsilon),
 \label{ac}
\end{equation}
\begin{equation}
\delta _\epsilon \chi _i  = (2/k) {\tilde {\cal D}} _{i} \epsilon ,
\label{bc}
\end{equation}
\begin{equation}
\delta _\epsilon A_i = -i \sqrt{2} (\bar\epsilon \chi _i -
\bar \chi _i \epsilon). \label {cc}
\end{equation}
The expression (\ref{a}) can be extended on the complex non-Majorana spinors
\begin{equation}
\delta _\epsilon \lambda =b^{-1} \epsilon + ib(\bar \epsilon
\gamma ^i \lambda) \partial_i \lambda . \label{nc}
\end{equation}
Now we have to use the above considered complex
supershift and superlightcone constraints of the Kerr geometry
(\ref{subman}, \ref{sshift}) in N=2 supergravity.
One should note, that the above superlightcone constraints restrict
values of supershift parameters $\bar\zeta$, leaving the values
of $\zeta$ free, which leads in general case to a non-Majorana spinor
\begin{equation}
\epsilon=
\left(\begin{array}{c}
 {\zeta}_{\alpha}\\
{\bar\zeta}^{\dot\alpha}
\end{array} \right)
\label{eps}\end{equation}
corresponding to a complex supershift.
Further, we should note that supercovariant derivative contains the
nonlinear  in $\epsilon$ terms from torsion ${\cal D} _N$, which hinders
the indirect use of finite supertranslations. At the other hand the Kerr
constraints on $\bar\zeta$, (\ref{sshift}), select a submanifold
displaying a remarkable nilpotency $\bar\zeta ^2=0$.
It means, that the values of $\bar\zeta$ on this submanifold lie in a
degenerate subalgebra of the Grassmann algebra \cite{finkim,Ber}.
\footnote{It means that the complex expansion
$  \zeta_{\alpha} =\Sigma \zeta^{(1)}_{\alpha i} s_q +
\Sigma \zeta ^{(2)}_{\alpha pqr} s_p s_q s_r ...., $
where $s_p s_q +s_q s_p =0 ,$
contains there only the first non-zero term
$s^2 =\bar s^2 = 0; \quad\bar s s+s \bar s =0$.
This degeneracy take place only on the considered subspace of
supershifts, however, even in a very narrow neighborhood of this subspace
the full algebra has to be taken into account. In particular, by application
spinor supercovariant derivatives when the Grassmann variables have to be
considered as independent.}
\par
To avoid the non-linear terms from torsion we will extend this property on
the four-component spinor $\epsilon$, and will restrict supershift by
the form
\begin{equation}
\epsilon=2^{-1/2}
\left(\begin{array}{c}
 {s \eta}_{\alpha}\\
{s \bar\zeta}^{\dot\alpha}
\end{array} \right)
\label{s-eps},\end{equation}
where
\begin{equation}
s^2 =\bar s^2 = 0; \quad\bar s s+s \bar s =0.
\end{equation}
This form possesses the nilpotency $\epsilon ^2= 0$
that leads to cancelling the non-linear term ${\cal D} _N$,
and gives rise to validity  (\ref{ac}), (\ref{bc}), (\ref{cc})
under finite supertranslations.
The formulated in N=1 case problem of the absorption of the Goldstone
fermion $\lambda$ is reduced to a choice of supergauge field
$$\epsilon(x)=-b\lambda(x),$$ that leads to satisfying (\ref{nc}) and
yields $\lambda ^\prime = \lambda + \delta \lambda \rightarrow 0, $
if we assume that $\lambda$ has corresponding nilpotency.
\par
Below we present few examples of supershift functions and will see that
the most interesting case is when the Goldstone fermion $\lambda$ satisfies
the massless Dirac equation.
\par
\section{Examples of N=2 super-geometries}
\par
In this section we consider different variants of the Goldstone fermion
satisfying the constraints (\ref{sshift}) and describe the resulting
super-geometries.
It is convenient to perform all calculations in the local Lorentz frame
of the Kerr geometry where  $\sigma$-matrices take the form
\cite{AG1,AG3,EinFink}
\begin{equation}
\sigma _i  =  2^{1/2}
\left( \begin{array}{cc}
e^3 _i& e^2 _i \\
e^1 _i& -e^4 _i
\end{array} \right);
\qquad
\bar\sigma _i  =  2^{1/2}
\left( \begin{array}{cc}
- e^4 _i& - e^2 _i \\
- e^1 _i& e^3 _i
\end{array} \right).
\label{sigi}
\end{equation}
The corresponding matrices with tetrad indices $a, b,c,d...$ will
be given by $\sigma ^a=\sigma _i e^{ai}$ and take the form
\begin{equation}
\sigma ^a  =  2^{1/2}
\left( \begin{array}{cc}
\delta _4^a& \delta _1^a \\
\delta _2^a& -\delta _3^a
\end{array} \right).
\label{siga}
\end{equation}
The four component spinor
\begin{equation}
\phi =
\left( \begin{array}{c}
A\\
B\\
C\\
D
\end{array} \right)
\end{equation}
will be called as aligned to $e^3$ if $e^3_i\gamma^i \phi^{(3)} = 0$ that
has the consequence $A=D=0$. Similarly, spinor $\phi ^{(4)}$ is aligned
to $e^4$ if $e^4_i\gamma^i \phi^{(4)} = 0$, which yields
 $B=C=0$. Therefore, any four component spinor can be
split on the sum of two aligned spinors
$\phi = \phi ^{(3)}+ \phi ^{(4)} $.
Physically, the $e^3 $ ($e^4 $) aligned solution describes the waves in
the $e^3 $ ($e^4 $) null direction.
In many cases (especially massless) solutions of these aligned parts are
independent. In the algebraically special Kerr-Newman geometry  all the
tensor fields are aligned to the principal null direction $e^3$.
The supershift constraints (\ref{sshift}) represent in fact a condition
for the two-component spinor $\bar\zeta$ to be aligned to $e^3$.
In the local Lorentz frame it takes the form $\bar\zeta ^{\dot 2} =0$,
the second component  $\bar\zeta ^{\dot 1 }=C $ is free.
Therefore, the general form of supershift satisfying the Kerr's
superconstraints and nilpotency condition will be
\begin{equation}
\epsilon= s 2^{-1/2}
\left(\begin{array}{c}
 {\eta}_{\alpha}\\
\bar\zeta ^{\dot \alpha}
\end{array} \right),
\label{nil-eps}\end{equation}
where
\begin{equation}
\eta_{\alpha} =
\left(\begin{array}{c}
 A\\
 B
\end{array} \right) ,
\label{eta}\end{equation}
and
\begin{equation}
\bar\zeta_{\dot\alpha} =
\left(\begin{array}{c}
 C\\
 0
\end{array} \right) .
\label{zeta}\end{equation}
For the sake of convenience we also give contravariant components
of $\eta $
\begin{equation}
\eta^{\alpha} =
\left(\begin{array}{c}
 B\\
 -A
\end{array} \right).
\end{equation}
The resulting complex Rarita-Schwinger field has contributions from metric
\begin{equation}
 \chi _g=\chi _{ga} e^a = (s\sqrt{2}/k)
\left(\begin{array}{c}
 dA +AH -B d\bar Y\\
d B -AG -BH\\
dC -C\bar H\\
C  dY
\end{array} \right),
\label{chig}\end{equation}
and  from electromagnetic field
\begin{equation}
 \chi _F= -i(s/\sqrt{2})
\left(\begin{array}{c}
 C \bar N e^3\\
C (\bar S e^3 - \bar N e^1)\\
A(N e^4 + S e^1) + B( N e^2 - S e^3)\\
-A N e^1 + BN e^3
\end{array} \right),
\label{chiF}\end{equation}
where we have introduced the notations for the geometry  parameters
\begin{equation}
H=[(\bar Z -Z)h - h,_4] e^3 /2; \quad G= h\bar Z e^1 -(h,_2-h Y,_3) e^3,
\label{GH}\end{equation}
and for the combinations of the tetrad components of electromagnetic field
\begin{equation}
N = F_{12} + F_{34},\quad S= 2 F_{31}.
\label{NS}\end{equation}
The function $Y(x)$ is the main function determining the principal null
congruence (\ref{k}) and the Kerr tetrad. It can be expressed as a
projective spinor coordinate $Y= \bar \psi^0 / \bar\psi ^1$
\footnote{This function is determined by the Kerr
theorem \cite{BKP,DKS}.
A fixation of $Y$ selects  the null planes in $CM^4$ and
null rays of the principal null congruence.}, and  satisfies the
equation
$dY = Z [e^1 - (P_{\bar Y}/P)e^3],$ where
$P= 2^{-1/2}(1+ Y \bar Y) $
for the stationary Kerr-Newman background.
The corresponding Kerr tetrad $e^a $ is  given in the Appendix C.
The resulting nilpotent contribution to tetrad contains two  terms,
contributions from metric and from electromagnetic field
$\delta e^a = \delta _g e^a + \delta _F e^a $.
The metric part will be
\begin{eqnarray}
\delta _g e^1 &=&
\bar s s \sqrt{2} [(\bar C C -\bar B B) dY - \bar A dB + Bd\bar A
+ \bar A A G + \bar A B (H+ \bar H)] ,
\nonumber \\
\delta _g e^2 &=&
- \delta \bar {e^1} ,
\nonumber \\
\delta _g e^3 &=&
\bar s s \sqrt{2} [(\bar A B) d\bar Y  -A \bar B dY +A d \bar A
 -\bar A dA + A\bar A(\bar H - H)] ,\nonumber \\
\delta _g e^4 &=&
\bar s s \sqrt{2} [\bar C dC -C d \bar C + \bar B dB -B d\bar B + \nonumber\\
&&\bar A B \bar G - A \bar B G  +(B \bar B- C \bar C )(\bar H - H)].
\label{egm}
\end{eqnarray}
Electromagnetic contribution to tetrad is
\begin{eqnarray}
\delta _F e^1&=&-ik\bar s s /2 [e^1(C\bar A \bar N - \bar C A N) -
C\bar A \bar S e^3], \nonumber \\
\delta _F e^2 &=& - \delta_F \bar {e^1}, \nonumber \\
\delta _F e^3 &=&i k\bar s s  e^3 ( C\bar A \bar N + \bar C A N)/2 ,
\nonumber\\
\delta _F e^4 &=&i k\bar s s  (N e^4 + S e^1) ( C\bar A  + \bar C A )/2.
\label{edF}
\end{eqnarray}
Nilpotent contribution to vector potential (\ref{cc})
also contains two terms $\delta A=\delta _g A_{a} e^a +\delta _F A_{a} e^a,$
\begin{eqnarray}
 \delta _g A=-i(\bar s s \sqrt{2}/k)(\bar C dA +\bar A dC -
A d \bar C - C d\bar A + \nonumber\\
2 A \bar C H - 2 \bar A C \bar H + 2 \bar B C dY - 2 B \bar C d \bar Y),
\label{dDA}\end{eqnarray}
and
\begin{eqnarray}
 \delta _F A=\bar s s /2 \lbrace e^1 [S A\bar A + (\bar N -N) A \bar B]
+ e^2 [\bar S A\bar A - (\bar N -N) \bar A  B] + \nonumber \\
e^3 [ (B\bar B + C\bar C) (N +\bar N) - A \bar B \bar S - \bar A B S ] +
e^4 A\bar A (N+ \bar N)\rbrace  .
\label{dFA}\end{eqnarray}
In these expressions we have the free spinor components
\begin{equation}
\eta_{\alpha} =
\left(\begin{array}{c}
 A(x)\\
 B(x)
\end{array} \right) ,
\label{eta1}\end{equation}
determining the supershift.
Now we consider some particular cases leading to
simplification of general expressions.
\par
\subsection{Case I}
\par
The simplest spinor shift  $A(x)\ne 0,\quad B=0,$ considered in the basis
having the $\sigma$-matrices  adapted to the auxiliary
Minkowski space $\eta_{ik}$.
The peculiarity of this shift is that while transformed to the aligned basis
\footnote{Transformations of the Dirac spinors from the $e^3$-aligned  basis
to the basis of auxiliary Minkowski space are given in Appendix D.}
this spinor  takes the same simple form that allows to simplify expressions
\begin{equation}
\eta_{\alpha} =
\left(\begin{array}{c}
 A(x)\\
 0
\end{array} \right) ,
\label{I}\end{equation}
We will also assume that $C=1,$ and obtain for $\delta_g$ contribution
to tetrad
\begin{eqnarray}
\delta _g e^1 &=&
\bar s s \sqrt{2} [ dY + \bar A A G ], \nonumber \\
\delta _g e^2 &=&
- \delta \bar {e^1}, \nonumber \\
\delta _g e^3 &=&
\bar s s \sqrt{2} [A d \bar A -\bar A dA + A\bar A(\bar H - H)], \nonumber \\
\delta _g e^4 &=&
\bar s s \sqrt{2} (\bar H - H),
\label{edg1}
\end{eqnarray}
and $\delta _F$ contribution
\begin{eqnarray}
\delta _F e^1 &=& -ik\bar s s /2 [e^1(\bar A \bar N -  A N) -
\bar A \bar S e^3], \nonumber \\
\delta _F e^2 &=& - \delta_F \bar {e^1}, \nonumber \\
\delta _F e^3 &=& i k\bar s s  e^3 ( \bar A \bar N + A N)/2, \nonumber \\
\delta _F e^4 &=& i k\bar s s  (N e^4 + S e^1) ( \bar A  + A )/2.
\label{edF1}
\end{eqnarray}
Contributions to vector potential $A$ will be
\begin{equation}
\delta _g A = -i(\bar s s \sqrt{2}/k)(dA - d\bar A + 2 A H -
2\bar A \bar H),
\label{dgA1}\end{equation}
\begin{equation}
 \delta _F A = \bar s s /2 [ (e^1 + e^2) A\bar A  + (e^3 + e^4 A\bar A)
(N+ \bar N)].
\label{dFA1}\end{equation}
One can see that the wave oscillations of the shift, $A = e^{ip_i x^i}$ can
lead to the wave oscillations of the terms $\delta _F e^a$ in tetrad and
the term $\delta _g A$ in the vector potential.
Since in the Kerr-Newman geometry function $ N$ has the behavior $N\sim Z^2$,
the tetrad oscillations are growing near the Kerr singular ring and take
the form of waves travelling along the singularity.
However, in this case we do not have clear proposals concerning
mechanisms or the possible origin of such oscillations in supershift.
Some suggestion to this case follows from observation that such oscillating
spinor $\eta$ resembles oscillating solutions of the Dirac equation.
\par
\subsection{Case II}
\par
The spinor shift has parameters  $A = 0,\quad B\ne 0,$ considered in the
basis where $\sigma$-matrices are adapted to the auxiliary  Minkowski
space $\eta_{ik}$.
Being transformed to the aligned basis this spinor  takes the form
$ A= B \bar Y; \quad B\ne 0$.  This case does not lead to essential
simplifications in respect to the general case, and we will not
write down the expressions for this case, but we should note that
appearance of traveling waves for oscillating $B$ can be observed in
this case, too.
\par
\subsection{ Case III}
\par
The case of supershifts aligned to $e^3$:
$A = 0;  B\ne 0,$ considered in the aligned $e^3$ basis.
\footnote{If $\sigma$-matrices were adapted to auxiliary Minkowski frame
this shift is given by parameters $ A= - \bar Y; \quad B=1.$ See matrices
of transformation in Appendix D.}
This case yields maximal simplifications of the expressions and is
motivated by super-QED model of broken supersymmetry for the region of
matter fields. Besides, as we shall see it leads to a family of
self-consistent super-Kerr-Newman solutions.
\par
There is no electromagnetic contributions to  tetrad in this case and we
have
\begin{eqnarray}
\delta e^1 &=&
\bar s s \sqrt{2} (C \bar C - B \bar B) dY , \nonumber \\
\delta e^2 &=&
- \delta \bar {e^1},\\
\delta e^3 &=& 0, \nonumber \\
\delta e^4 &=&
\bar s s \sqrt{2}  [(C \bar C - B \bar B) h(\bar Z - Z) e^3+
\bar C dC -C d \bar C + \bar B dB -B d\bar B]. \nonumber
\label{edg3}
\end{eqnarray}
The contribution to the Kerr-Newman vector potential will be
\begin{equation}
\delta _g A = -i(\bar s s 2\sqrt{2}/k)[\bar B C dY  - B \bar C d \bar Y],
\label{dgA3}\end{equation}
\begin{equation}
 \delta _F A = -\bar s s /2 e^3 [(B\bar B + C \bar C)(N+ \bar N)] .
\label{dFA3}\end{equation}
The case $C=\bar B$ could be called pseudo-Majorana. In this case there are
no nilpotent contributions to tetrad at all, $\delta e^a = 0$, and metric
has pure bosonic form similar to the case of extreme black holes without an
angular momentum. However, the Kerr-Newman vector potential has nilpotent
contributions, and moreover, it can have traveling waves from the term
\begin{equation}
\delta _g A=-i(\bar s s 2\sqrt{2}/k)(\bar B ^2 dY  - B^2 d \bar Y).
\label{dgA4}\end{equation}
In the case of independent $B$ and $C$ there also appears tetrad traveling
waves.
\par
In general case there is a non-zero torsion in the super-Kerr geometry
${Q}^a = \bar\psi _b e^b \wedge \gamma ^a \psi _c e^c $ which
gives rise to corresponding traveling waves of torsion.
\footnote{The Kerr-Newman solution with torsion based on the
Poincar\'e gauge theory has been considered in \cite{tors}.}
\par
The torsion $ {Q}^a$ contains contributions from metric $Q^a_g$,
electromagnetic field $Q^a_F$ and terms of interplay $Q^a_I$.
\par
We write down here the torsion terms only for the simplest and the most
interesting case III, when $A=0$.
Contributions from metric are
\begin{eqnarray}
Q^1_g &=& i \bar s s \sqrt{2}/ k^2 ( C d \bar C - C\bar C H
- \bar B dB + \bar B B H) \wedge dY, \nonumber\\
Q^2_g &=& \bar Q^1_g ,
\nonumber\\
Q^3_g &=& i \bar s s \sqrt{2}/ k^2 (C \bar C - B \bar B)dY \wedge d \bar Y,\\
\nonumber
Q^4_g &=& i \bar s s \sqrt{2}/ k^2 [ (d \bar C - \bar C H)\wedge
(dC - C \bar H)+ (d \bar B - \bar B \bar H) \wedge (dB - BH)].
\label{Qg}
\end{eqnarray}
Contributions from electromagnetic field are
\begin{eqnarray}
Q^1_F &=& i \bar s s /(2\sqrt{2}) N \bar N
(B \bar B - C \bar C) e^1 \wedge e^3 ,
\nonumber\\
Q^2_F &=& \bar Q^1_F ,
\nonumber\\
Q^3_F &=& 0 ,
\nonumber\\
Q^4_F &=& i \bar s s /(2\sqrt{2}) (B \bar B - C \bar C)
(\bar N e^1 - \bar S e^3)\wedge (N e^2 -S e^3).
\label{QF}
\end{eqnarray}
Terms of interplay are
\begin{eqnarray}
Q^1_I &=& \bar s s /(k\sqrt{2})N(B d \bar C - \bar C d B)\wedge e^3,
\nonumber\\
Q^2_I &=& \bar Q^1_I , \\
Q^3_I &=& 0 , \nonumber \\
Q^4_I &=& \bar s s /(k\sqrt{2})[B(d\bar C -\bar C H) \wedge (N e^2 - S e^3)
- C(d\bar B -\bar B \bar H) \wedge (\bar N e^1 - \bar S e^3)\nonumber \\
+ c.c.term ]. \nonumber
\label{QI}
\end{eqnarray}
In the case $\bar B =C$ all the torsion terms disappear.
\par
\noindent
\section{Self-consistent super-Kerr-Newman solutions to broken N=2
supergravity}
\par
\noindent
The generalized to N=2 supergravity Deser-Zumino Lagrangian
(\ref{L}),(\ref{Lsg}) takes the form
\begin{equation}
{\cal L} = - ( i / 2 ) [ \bar \lambda \gamma {\tilde {\cal D}} \lambda
- \bar {\tilde {\cal D}} \bar \lambda \gamma \lambda ]
- ( i / 2 b )[\bar \lambda \gamma ^i \chi _i
-\bar \chi _i \gamma ^i \lambda ]
+ {\cal L} _{2-sg},
\label{LN2}\end{equation}
where the N=2 supergravity Lagrangian is
\begin{equation}
{\cal L} _{2-sg}= -eR/2k^2 - 1/4 F_{ij} F^{ij} -
i \epsilon^{ijkl} \bar \chi _i \gamma _5
\gamma _j {\tilde {\cal D}} _k \chi _l.
\label{Lsg-2}\end{equation}
\par
It follows from (\ref{LN2}) that the self-consistent solutions to broken
N=2 supergravity has to take into account the energy-momentum tensor of the
Grassmann fields.
In particular, when considering the initiate {\it trivial} solutions in
the super-gauge with zero Rarita--Schwinger field, one can use this
Lagrangian with $\chi=0$ that yields the Einstein--Maxwell--Dirac system
of equations. We note that the energy-momentum tensor of the Goldstone
field $\lambda$ acts here as fermionic matter.
However, when using the {\it exact} Kerr-Newman solution as trivial one
to perform the super-gauge with absorption of the Goldstone fermion,
we do not take into account the energy-momentum
tensor of the Goldstone field. Therefore, in general case, the considered
above super-geometries cannot be treated as self-consistent.
However, one exclusive case can be selected when the self-consistency
is guaranteed. It takes place for the ghost Goldstone field possessing
the zero energy-momentum tensor.
\par
In this case, starting from the Lagrangian with $\chi =0$, we have in fact
the Einstein-Maxwell system of equation leading to the exact Kerr-Newman
solution and the Dirac equation ( on the Kerr-Newman background ) for
the Goldstone fermion $\lambda$.
\par
This solution can be considered as an exact super-solution to N=2
supergravity coupled to Goldstone field. Then, absorption of the
Goldstone field by the complex Rarita-Schwinger field $\chi$ turns this
solution into self-consistent solution with broken N=2 supersymmetry.
\par
 In the Appendix B we have given the  solutions of the massless Dirac
equation on the Kerr-Newman background in the aligned to $e^3$ case
(case III).
The corresponding functions $B$ and $C$ have the form
\begin{equation}
B= \bar Z f_B (\bar Y, \bar \tau)/P,\qquad
C=  Z f_C ( Y,  \tau)/P,\qquad
\label{BC}\end{equation}
where $f_B$ and $f_C$ are arbitrary analytic functions of
the complex angular variable
\begin{equation}
Y=e^{i\phi} \tan (\theta/2),
\label{Y}\end{equation}
and the retarded-time
\begin{equation}
\tau = t - r -ia \cos \theta
\label{tau}\end{equation}
satisfies the relations
$\tau,_2=\tau,_4 =0, $ and $Y,_2=Y,_4=0$.
\par
The energy-momentum tensor of the Goldstone field $\lambda$ can be expressed
via the Grassmann contributions to tetrad (\ref{edg3}) as follows
\begin{equation}
T_{ik} = i/2 e_{(ia} \delta e^a _{k)}.
\label{T}\end{equation}
One sees that the energy-momentum tensor of the Goldstone field $\lambda$
with $C=\bar B$ cancels, $T_{ik} = 0$, and the field takes the ghost
character.
As it was mentioned earlier, in this case torsion and Grassmann
contributions to tetrad are absent, and metric takes the exact
Kerr-Newman form.
\par
The main features of the resulting super-Kerr-Newman solutions are the extra
wave fields on the bosonic Kerr-Newman background:
the complex Rarita-Schwinger field $\chi _i$ and the nilpotent contributions
to electromagnetic field given by the expressions (\ref{dFA3}) and
(\ref{dgA4}).
\par
One should note that the expressions for $B$ and $C$ are singular on the
Kerr singular ring,
$Z^{-1}\equiv P^{-1}(r+ia \cos \theta )=0$, and
contain traveling waves if there is an  oscillating dependence on
complex time parameter $\tau$.
Indeed, near the Kerr singular ring $\tan \theta \simeq 1$, and
angular dependence of these solutions on $\phi$ is determined by the
degree of function $Y=e^{i\phi} \tan (\theta/2)$. One sees that any
non-trivial analytic dependence on $Y$ will lead to a singularity in
$\theta$. Thus, besides the Kerr singular ring the solutions contain an
extra axial singularity which is coupled topologically with singular
ring threading it. The singular Rarita-Schwinger `hair'
was mentioned earlier by Aichelburg and G\"uven \cite{AG3,Aich2} as an
obstacle to form a super black hole in the case of the Kerr geometry.
However, when considering this solution as a model of
spinning particle, we have to treat a region of
parameters $a\gg m $ leading to a naked disk-like super-source
\cite{caris,bls,super-K} without horizons that allows to avoid the
objections related to rotating black holes.
\footnote{Corresponding fermionic singular solutions were also considered in
\cite{EinFink}}.
However, even in the case of black hole, one can assume that this `axial'
singularity will not hinder to form horizon since it is build of the ghost
fields.
\par
One should also note that this `axial' singularity in some solutions can
change the position in time scanning the space-time.
One simple example of such function is
$$C= (Z/P)[ f_1(\tau) - Y f_2 (\tau)]^{-1},$$ where the position of the
`axial' singularity is determined by the root of equation $Y= f_1/f_2 $
depending on the retarded-time parameter $\tau$.
 \footnote{Topological coupling of the axial singularity with the Kerr
singular ring provides its guiding role for the ring that allows to consider
a speculation that this couple of singularities could play the role of a
topological wave-pilot construction in the spirit of the old de Broglie
ideas.}
\par
Let us consider elementary fermionic wave excitations in the form
\begin{equation}
C= \bar B = Z/P Y^n e^{ i\omega \tau }=1/(r+ia\cos \theta )
\tan ^n (\theta /2).
\label{el} \end{equation}
Taking into account the coordinate relation of the Kerr geometry
\begin{equation}
 x+iy = (r+ia) e^{i\phi} \sin \theta,\qquad z=r \cos \theta,
\label{xyz}\end{equation}
one sees that near the Kerr singular ring
 (where  $r\simeq 0, \quad\sin \theta \simeq 1$ ) at the point $\phi =0$,
singularity is directed along the $x$-axis.
One can introduce the local coordinates $\tilde z=z$ and $\tilde y = y-a$
on the plane ortogonal to direction of singularity. Then, near the
singularity
$\tilde r \simeq \sqrt {2a(\tilde y + i  \tilde z)}$, and
 (\ref{el}) takes the form
\begin{equation}
C \simeq {1 \over \sqrt {2a(\tilde y + i  \tilde z)}}
e^{i(n\phi + \omega t)},
\label{simel}\end{equation}
that describes a traveling wave along the ring-like singularity parametrized
by  $\phi$.
The first factor shows that this singularity is a branch line corresponding
to the known twofoldedness of the Kerr geometry.
Because of that parameter $n$ can take both integer and half-integer values.
The potential (\ref{dgA4}) takes near singular ring the form
\begin{equation}
\delta _g A \simeq (\bar s s/k) \{
{[a(\tilde y + i \tilde z)]}^{-3/2} e^{i(2n\phi + 2\omega t)}
( d\tilde y+id\tilde z) +  c.c.term \}.
\label{simAg}\end{equation}
The obtained from potential (\ref{dgA4}) electromagnetic field is
\begin{equation}
F= -i(\bar s s 2^{5/2}/k) \{
C^2 [ \tilde r ^{-1} dY \wedge dt + \tilde r ^{-1}
(\tilde r ^{-1} -i \omega ) e^1 \wedge e^3 ] - c.c.term \}.
\label{Fg}\end{equation}
The proportional to $ e^1 \wedge e^3 $ term represents the null
electromagnetic field propagating along principal null direction $e^3$.
Near the Kerr singular ring this congruence is tangent to ring leading to
traveling waves propagating along the ring.
\par
The `axial' singularity coincides with z-axis and can be placed either at
$\theta=0\quad (Y=0)$ or at $\theta=\pi \quad (Y=\infty)$. It is
a half-infinite line threading the Kerr singular ring and passing to
`negative' sheet of the Kerr geometry.
Its position and character depend on the values of $n$.
By introducing the distance from `axial' singularity $\rho= \sqrt{x^2+y^2}$,
one can describe its behavior in the asymptotic region of large $r$ by the
following expressions:
\par
- if $\theta \simeq 0$ then
$$\delta_g A \sim \rho^{2n} r^{-3-2n}(dx+idy),  $$
$$\delta_F A \sim \rho^{2n} r^{-4-2n}(dz+dt), $$
\par
- if $\theta \simeq \pi$ then
$$\delta_g A \sim \rho^{-2n-2} r^{2n-1}(dx-idy),  $$
$$\delta_F A \sim \rho^{-2n-2} r^{2n-2}(dz-dt). $$
One sees that this singularity can be increasing or decreasing function of
distance $r$. For some $n$ (for example n=1/2,-3/2) dependence on $r$ can
disappear. The solutions with `increasing' and `even' singularities cannot
be stable.
In the cases $n=0$ and $n=-1$ singularity represents a `decreasing'
half-infinite line like the string of the Dirac monopole.
The case $n=-1/2$ is exclusive: there are two `decreasing' singularities
which are situated symmetrically at $\theta=0$ and  $\theta=\pi$.
The space part of the null vector $e^3$ is tangent to axial singularity,
and electromagnetic field (\ref{Fg}) grows near this singularity and
contains in asymptotic region the leading term in the form of the
null traveling wave
\begin{equation}
F\simeq -(\bar s s 2^{5/2}/k) [ C^2 \tilde r ^{-1} \omega e^1 \wedge e^3
+ c.c.term ].
\label{Fgass}\end{equation}
\par
\section{Conclusion}
\par
We have shown that the problem of obtaining non-trivial super black hole
solutions can be resolved in supergravity broken by Goldstone
fermion. In case of the Kerr geometry it leads to a specific non-linear
realization of supersymmetry which has to be adjusted with complex structure
of the Kerr geometry.
\par
We considered three families of the rotating and charged super black hole
geometries representing supergeneralizations of the Kerr-Newman geometry
generated by Goldstone fermions of various form.
When supersymmetry is broken, the Goldstone fermion disappears and these
super-geometries can acquire torsion and traveling waves of the
Rarita-Schwinger and electromagnetic fields.
\par
Among these geometries we have selected one exclusive family of the
{\it exact} super-Kerr-Newman solutions to N=2 supergravity broken by
Goldstone fermion aligned to principal null congruence of the Kerr
geometry.
\par
Peculiarities of these solutions were analyzed, in particular,
the ghost character of Goldstone field, absence of torsion, the appearance
of an extra `axial' singular line which is coupled topologically to the
Kerr singular ring, and traveling waves of the null electromagnetic
field concentrated near the `axial' and ring-like singularities.
\par
  The obtained exact solutions are based on the massless Goldstone field.
At present stage of investigation our knowledge regarding the origin
of the Goldstone fermion is very incompleted. Analyzing the Wess-Zumino
model of super-QED \cite{WB} with spontaneously broken supersymmetry one
can see that it leads to massless Goldstone fermions, at least in the
region that is out of core. It takes the place if we are interested
mainly in the stationary black hole solutions describing the region of
massless fields.
\par
However, as it was mentioned in \cite{super-K,superBH}, for the known
parameters of spinning particles the angular momentum is very high,
regarding the mass parameter. In this case the black hole horizons
disappear and a `hard core' region, representing a superconducting
disk-like source, has to be taken into account. Treatment of this
region is extremely important and complicated problem, however, it is
beyond of the frame of this paper.
We should only note that investigations of such super-sources should
be connected apparently with Seiberg-Witten theory \cite{SW} and with
some other interesting models of broken supersymmetry together with
gauge symmetry breaking \cite{APT,FGP,BG,IZ}.
\par
\section*{Acknowledgments}
\par
I am grateful to P. Aichelburg, F. Hehl,  G. Gibbons and J. Maharana
for interest to this work and stimulating discussions, and also to
B.G. Sidharth for useful discussions and very kind hospitality at B.M. Birla
Science Centre ( Hyderabad) where this work was finished.
\par
\section*{ Appendix A. Some relations of DKS-formalism \cite{DKS} }
The rotation coefficients can be obtained from the following independent
forms
\begin{eqnarray}
\Gamma_{42} = - Z e^1 - Y,_3 e^3; \nonumber \\
 \Gamma_{12} + \Gamma_{34} = [h,_4 + (\bar Z -Z) h] e^3 ; \nonumber \\
\Gamma_{31} = h Z e^2 + (h,_1 - h \bar Y,_3) e^3;
\end{eqnarray}
taking into account  property of skew-symmetry $ \Gamma_{ab}= - \Gamma_{ba}$
and relations to the complex conjugate forms, for example:
$ \Gamma_{12}= \bar \Gamma_{21} $;
$ \Gamma_{42} = \bar \Gamma_{41}$;
$ \Gamma_{32} = \bar \Gamma_{31}. $
\par
Basic relations:
\begin{eqnarray}
P&=&2^{-1/2}(1+Y \bar Y);\qquad Y=e^{i\phi} \tan {\theta \over 2};\\
 Z,_4 &=&-Z^2;\qquad \bar Z,_4 =-\bar Z^2; \qquad Z,_2 = (Z-\bar Z)Y,_3; \\
 Y,_1 &=&Z;\qquad Y,_2 = Y,_4 = 0;
\qquad \bar Y,_2 =\bar Z;\qquad \bar Y,_1 = \bar Y,_4 = 0;\\
 Y,_3 &=&- Z P_{\bar Y} /P= -Z \bar Z^{-1} \bar Z P,_2/P .
\label{A1}
\end{eqnarray}
Some other useful relations:
\begin{eqnarray}
P_Y P_{\bar Y}/P^2 &=& {1 \over 4} \sin ^2 \theta; \\
P_Y - r,_1 &=& - ia(\cos \theta),_1 = -iaZ 2^{1/2} P_Y/P^2; \\
(r,_1 - P_Y )e^1 + (r,_2 - P_{\bar Y})e^2 &=&
ia 2^{1/2}/P^2 [P_{\bar Y}  d \bar Y - P_{Y} dY]=  \nonumber \\
4a P_Y P_{\bar Y}/P^2 d\phi &=& a\sin^2 \theta d\phi ;
\end{eqnarray}
\begin{eqnarray}
(r,_1 - P_Y )(r,_2 - P_{\bar Y} ) &=& 2a^2 P_{Y} P_{\bar Y} Z \bar Z / P^4;
\nonumber\\
&=& a^2 \sin^2 \theta / (r^2+ a^2 \cos^2 \theta);
\end{eqnarray}
\begin{equation}
dY \wedge d\tau = \tilde r ^{-1} e^1 \wedge e^3.
\label{null}\end{equation}
The complex radial coordinate $\tilde r =  r+ ia\cos \theta = PZ^{-1}$.
\par
Some tetrad derivatives
\begin{eqnarray}
Z,_1  &=& (Z^3 /P) F''_{YY} + 2  Z^2  P_{Y} /P , \\
{\bar Z},_2  &=& (\bar Z^3 /P) \bar F''_{YY} + 2 \bar Z^2  P_{\bar Y}/P ,\\
(\ln \bar Z /P),_2 &=& {\bar Z}^2 2A/P  - Y,_3 \bar Z / Z , \\
Z,_3 /Z &=& - (Z,_1 /Z) P_{\bar Y} /P +hZ -
Z P_{Y \bar Y}/P + (Z- \bar Z) P_Y P_{\bar Y}/P^2, \\
\tilde r,_2 &=& P_{\bar Y}, \qquad \tilde r,_4 = P .
\end{eqnarray}
\par
We should note that in definitions of DKS-paper the "in-going" congruence
is used leading to the "advanced" time coordinate
$\tau=\tau_{adv}= t + \tilde r _{adv} = t+r+ia\cos \theta$.\footnote{The
sign of $\rho $ in the expresssion (7.9c) of DKS is apparently wrong.}
In this case the "out"-congruence takes place on  the  "negative" sheet of
metric where $r\leq 0$. Using the redefinition
$\tilde r \rightarrow - \tilde r$  one can interchange the "positive"
and "negative" sheets of the Kerr geometry that yields the retarded time
$\tau_{ret}= t - \tilde r_{ret} = t-r-ia\cos \theta$
for positive values of $r$. Since this redefinition can be performed in
the final expressions we retain the DKS-notations in this paper for the
sake of convenience with the exceptions of the expressions for $\tau$
when we would like to underline that it is the retarded-time coordinate.
\par
\section*{Appendix B. Aligned solutions of the Dirac equation}
\par
The Dirac spinor is represented in general form
\begin{equation}
\Psi_D =
\left( \begin{array}{c}
A\\
B\\
C\\
D
\end{array} \right).
\end{equation}
The supercovariant derivative in complex N=2 case has the form
\begin{equation}
{\tilde {\cal D}} _{i} = {\cal D} _i +{\cal D} _{Ni} - (ik/2\sqrt{2}) F_{ab}
\Sigma ^{ab}\gamma _i.
\end{equation}
The metric part of the supercovariant derivative is
\par
${\cal D}_i = \partial_i - \frac{1}{2} \Gamma_{abi} \Sigma^{ab}.$
\par
For the Ricci rotation coefficients we use notations of DKS-paper \cite{DKS}
$\omega_{i a b} =- \Gamma_{abi}, $ see Appendix A.
\begin{eqnarray}
\frac{1}{2}\Gamma_{bc}\Sigma ^{bc} \Psi_D &=& \Gamma_{12} \Sigma^{12} \Psi_D
+ \Gamma_{31} \Sigma^{31} \Psi_D
+ \Gamma_{41} \Sigma^{41} \Psi_D \nonumber \\
&&+ \Gamma_{42} \Sigma^{42} \Psi_D + \Gamma_{34} \Sigma^{34} \Psi_D
+ \Gamma_{32} \Sigma^{32} \Psi_D.
\label{gamma}\end{eqnarray}
The Dirac equation $ \gamma^i {\tilde {\cal D}}_{i} \Psi_D = 0 $.
\par
A special class of $e^3$-aligned solutions satisfying the constraint
 $ \gamma^i e^3_i \Psi_D = 0 $,
has the form
\begin{equation}
\Psi_D =
\left( \begin{array}{c}
0\\
B\\
C\\
0
\end{array} \right).
\end{equation}
For the $e^3$-aligned solutions of the Dirac equation the Kerr-Newman
electromagnetic field drops out.
In the nilpotent case (\ref{nil-eps}) the nonlinear term drops out too.
The spinor-valued  1-form expression ${\cal D} \Psi_D $ takes the form
\begin{equation}
{\cal D} \Psi_D =
\left( \begin{array}{c}
-B (\bar Z e^2 + \bar Y,_3 e^3)\\
dB + B[h,_4 +(Z-\bar Z) h/2] e^3\\
dC + C[h,_4 -(Z-\bar Z) h/2] e^3\\
C ( Z e^1 +  Y,_3 e^3)
\end{array} \right).
\end{equation}
As a result, the Dirac equation
$ \gamma^a {\cal D} _a \Psi_D = 0 $ yields the four equations:
\begin{equation}
C,_4 + ZC =0;\qquad C,_2-C Y,_3 =0,       \label{C}
\end{equation}
and
\begin{equation}
B,_4 + \bar Z B =0;\qquad B,_1-B \bar Y,_3 =0.
\label{B}
\end{equation}
These equations can be easily solved by  using  the known basis relations
of the Kerr-Schild formalism (See Appendix A).
\par
First equation (\ref{C}) gives $C=Z C_0$ where $C_0,_4=0$.
Then, substituting $C$ and into second equation
we obtain  $ C=f Z/P$, where $f$ must satisfy the conditions
$f,_2=f,_4=0$.
Therefore, $f$ must be a function taking constant values on the
null planes spanned by vectors $e^1$ and $e^3$. These null planes are
"left" null planes of a foliation of space-time into  complex null
cones. In another terminology they represent a geometrical image of
twistors and can be parametrized by twistor coordinates.
All the twistor coordinates $\chi$ and
$\mu=x^i \sigma_i \chi$ satisfy the relations $(...),_2=(...),_4 =0$,
see \cite{BKP}.
Coordinate $Y$ is in fact one of the (projective) twistor coordinates.
For our problem it will be convenient to use a retarded time coordinate
$\tau$. The retarded time coordinate to a point $x$ is defined by a point
of intersection of the light cone emanated from $x$ with the world-line of
source.
\par
The light cone is split on the "left" and "right" null planes,
therefore, the retarded time parameter takes constant values on the
"left" null planes of the cone that leads to the relations
$\tau,_2=\tau,_4 =0$.
\footnote{Details of this construction can be found in \cite{BKP,c-str}.}
For the Kerr geometry one can use a known complex interpretation containing
a complex world line for the source. \footnote{One should note that in the
case of the complex world line the "left" and "right" roots for the retarded
time $\tau$ are different, and "right" root satisfies the relations
$\tau,_1=\tau,_4 =0$.}
In the rest frame the complex light cone equation $(t-\tau)^2
= \tilde r ^2$ can be split with selection of the retarded fold
$\tilde r = t -\tau$. Here $t$ is a real time coordinate,
and $\tilde r = r +i a \cos \theta = P/Z$ is a complex radial distance from
the real point $x$ to a point of source at the complex world line.
It yields $\tau = t - \tilde r$.
Therefore, the function $f$ can be an arbitrary analytic function of
complex coordinates $Y$ and $\tau$, and we have  solution
 $C=f(Y,\tau)Z/P$.
For function $B$ one can obtain similarly
 $B=\tilde f (\bar Y,\bar\tau) \bar Z /P$, where
 coordinates $\bar Y$ and $\bar\tau$ satisfy the relations
 $(...),_1=(...),_4 =0$ , and are constant on the "right" null planes
of the light cone foliation.
\par
\section*{ Appendix C. The Kerr tetrad and some useful matrix expressions }
The Kerr tetrad $e^a $ is determined by function $Y(x)$:
\begin{eqnarray}
e^1 &=& d \zeta - Y dv, \qquad  e^2 = d \bar\zeta -  \bar Y dv, \nonumber \\
e^3 &=&du + \bar Y d \zeta + Y d \bar\zeta - Y \bar Y dv, \nonumber\\
e^4 &=&dv + h e^3,\label{Kt}
\end{eqnarray}
where the null Cartesian coordinates are used
$\sqrt{2}u=z+t,\quad\sqrt{2}v=z-t,
\quad \sqrt{2}u=x+iy,\quad\sqrt{2}v=x-iy$.
\par
$\sigma$-matrices in the local Kerr geometry are
\begin{equation}
\sigma _i  =  2^{1/2}
\left( \begin{array}{cc}
e^3 _i& e^2 _i \\
e^1 _i& -e^4 _i
\end{array} \right) ;
\qquad
\bar\sigma _i  =  2^{1/2}
\left( \begin{array}{cc}
- e^4 _i& - e^2 _i \\
- e^1 _i& e^3 _i
\end{array} \right) .
\end{equation}
The corresponding matrices with tetrad indices $a, b,c,d...$ will
be given by $\sigma ^a=\sigma _i e^{ai}$ and take the form
\begin{equation}
\sigma ^a  =  2^{1/2}
\left( \begin{array}{cc}
\delta _4^a& \delta _1^a \\
\delta _2^a& -\delta _3^a
\end{array} \right) .
\end{equation}
\begin{equation}
\Sigma^{ab}= \frac{1}{4} (\gamma^a \gamma^b - \gamma^b \gamma ^a)
=\frac{1}{4}
\left( \begin{array}{cc}
\sigma^a \bar\sigma^b - \sigma^b \bar\sigma^a & 0  \\
0&\bar\sigma^a \sigma^b - \bar\sigma^b \sigma^a
\end{array} \right).
\end{equation}
In particular
\begin{equation}
\begin{array}{ccc}
\Sigma ^{12} & = & \frac{1}{2}
\left( \begin{array}{cccc}
-1&0&0&0 \\
0&1&0&0 \\
0&0&-1&0 \\
0&0&0&1
\end{array} \right) ;
\qquad
\Sigma ^{14} =
\left( \begin{array}{cccc}
0&1&0&0 \\
0&0&0&0 \\
0&0&0&0 \\
0&0&0&0
\end{array} \right) ;\\
\Sigma ^{24} & = & \left(
\begin{array}{cccc}
0&0&0&0 \\
0&0&0&0 \\
0&0&0&0 \\
0&0&-1&0
\end{array} \right) ;
\qquad
\Sigma ^{31} =
\left( \begin{array}{cccc}
0&0&0&0 \\
0&0&0&0 \\
0&0&0&-1 \\
0&0&0&0
\end{array} \right) ; \\
\Sigma ^{32} & = & \left(
\begin{array}{cccc}
0&0&0&0 \\
1&0&0&0 \\
0&0&0&0 \\
0&0&0&0
\end{array} \right);
\qquad
\Sigma ^{34} = \frac{1}{2}
\left( \begin{array}{cccc}
1&0&0&0 \\
0&-1&0&0 \\
0&0&-1&0 \\
0&0&0&1
\end{array} \right).
\end{array}
\end{equation}
\par
\section*{ Appendix D. Transformations of the Dirac spinors from the
$e^3$-aligned  basis to the basis of auxiliary Minkowski space}
\par
 Transformations of the Dirac spinors from the $e^3$-aligned  basis to
the basis of auxiliary Minkowski space are given by matrices
\begin{equation}
\left( \begin{array}{cc}
M^{-1}& 0 \\
0 & M^\dagger
\end{array} \right)
\left( \begin{array}{c}
\phi _{al}\\
\bar \chi _{al}
\end{array} \right)
=\left( \begin{array}{c}
\phi _{Mink}\\
\bar \chi _{Mink}
\end{array} \right),
\end{equation}
where
\begin{equation}
M =
\left( \begin{array}{cc}
1& \bar Y \\
0& 1
\end{array} \right) ,
\end{equation}
\begin{equation}
M^{-1} =
\left( \begin{array}{cc}
1& -\bar Y \\
0& 1
\end{array} \right) ,
\end{equation}
\begin{equation}
M^\star =
\left( \begin{array}{cc}
1& Y \\
0& 1
\end{array} \right),
\end{equation}
\begin{equation}
M^\dagger =
\left( \begin{array}{cc}
1& 0 \\
Y& 1
\end{array} \right).
\end{equation}
\par
\pagebreak

\end{document}